\documentclass[preprint]{elsarticle}
\usepackage{graphicx}
\usepackage{natbib}
\usepackage{dcolumn}
\usepackage{bm}
\usepackage{color}
\usepackage{epstopdf}
\DeclareGraphicsExtensions{.pdf,.eps}
\newcommand{\ket}[1]{\left| #1 \right>}

\newcommand{\meanval}[1]{\left< #1\right>}
\begin{document}
\title{Implementation of quantum logic gates using coupled Bose-Einstein condensates}
\author{F.~S. Luiz}
\address{Departamento de F\'{\i}sica, Universidade Federal de S\~ao Carlos. S\~ao Carlos-SP, Brazil}
\author{E. I. Duzzioni}
\address{Departamento de F\'{\i}sica, Universidade Federal de Santa Catarina, 88040-900, Florian\'opolis, SC, Brazil}
\author{L. Sanz}
\ead{lsanz@infis.ufu.br}
\address{Instituto de F\'{\i}sica, Universidade Federal de
Uberl\^andia, 38400-902, Uberl\^andia-MG, Brazil}
\begin{abstract}
In this work, we are interested on the implementation of single-qubit gates on coupled Bose-Einstein condensates (BECs). The system, a feasible candidate for a qubit, consists on condensed atoms of different hyperfine levels coupled by a two-photon transition. It is well established that the dynamics of coupled BECs can be described by the two-mode Hamiltonian which takes into account the confinement potential of the trap and the effects of collisions associated with each condensate. Other effects, as collisions between atoms belonging to different BECs and detuning are included in this approach. We demonstrate how to implement two types of quantum logic gates: \textit{transfer-population} gates (NOT, $\hat{Y}$, and Hadamard), which require inversion of population between hyperfine levels, and \textit{phase} gates ($\hat{Z}$, $\hat{S}$, and $\hat{T}$), which require self-trapping. We also discuss the experimental feasibility, checking the robustness of quantum gates against variations of physical parameters out of the ideal conditions for implementation of each quantum logic gate.
\end{abstract}
\maketitle

Quantum information processing (QIP) is one of the remarkable research topics in Physics during the last decades~\cite{Nielsen00,Ladd10}. Polinomial versus superpolinomial times in factorization and faster search algorithms~\cite{Shor94,Grover97} are two examples of the advantages of this kind of processing. The first requirement for the implementation of QIP is the definition of a quantum of information or qubit~\cite{Schumacher95}, which can be done using a two-level physical system. In addition to associate the logic states 0 and 1 with one of two levels, the dynamics of the system should allow the preparation of superpositions of these states, which must be robust against decoherence. Several systems has been pointed out as good candidates for the implementation of a qubit. Those include photons, ions, charge and spin in semiconductor nanostructures, Josephson junctions of superconductors, nuclear spins, and trapped neutral atoms~\cite{Ladd10}.

Concerning trapped atoms, the Bose-Einstein condensates (BECs) with alkalii atoms, created in 1995~\cite{MITBEC,JILABEC}, has been pointed out as promising candidates for implementation of QIP proposals because the long decoherence time scale and the high degree of manipulation and control of its physical properties. In particular, coupled BECs can be implemented considering two experimental setups. In the first one, atoms condensed in two different sites of an optical lattice are coupled by tunneling~\cite{Cataliotti01,Salgueiro07}.  The second configuration consists of condensed atoms in different hyperfine levels coupled by laser~\cite{Hall98a,Hall98b}. The two-mode model provides an adequate theoretical description of the dynamics of both systems~\cite{Milburn97}. Experimental achievements on coupled BECs include the observation of coherent oscillations~\cite{Hall98b,Cataliotti01,Albiez05}, the realization of coupled condensates inside a chip~\cite{Hansel01,Bohi2009}, the creation of atomic interferometers~\cite{Oberthaler10}, and entangled quantum states~\cite{Oberthaler08}. Because the process of transfer of population are performed with high quality in a full controlled experimental setup, the system can be considered, in first approximation, in an equilibrium state. Although, the effect of the presence of non-condensed atoms over the coupled condensate are one of the causes behind of the destruction of condensate, the time scales are long enough for QIP operations.

There are few theoretical proposals for the implementation of gates in coupled BECs. Those include the works of Calarco \textit{et al.}~\cite{Calarco00,Calarco02}, focussed on the analysis of the conditions for the implementation of a phase gate by manipulation of confinement potentials and a quantum logic gates using marker atoms~\cite{Calarco04}. Pachos and Knight~\cite{Pachos03} proposed two-qubits operations including a Toffoli gate by combining both, tunneling and adiabatic passage of condensate atoms in a superlattice. Lee \textit{et al.} explore the dipole-dipole interaction as a coupling mechanism~\cite{Lee05}. In a recent works, Byrnes \textit{et al.}~\cite{Byrnes15,Byrnes12} propose the codification of a quantum bit using collective states of coupled BECs by using an atomic coherent state, as well as the implementation of the two-qubit gate CNOT and an application to Grover algorithm. The authors point out that this proposal brings a natural advantage once the energy scale of the interaction is boosted by a factor corresponding to the number of condensed atoms, with the consequence of reducing the time of CNOT gate by the same factor. Concerning single-qubit operations, it is discussed how to rotate the qubit around $z$.

Nevertheless, the practical implementation of single-qubit quantum logic gates was not sufficiently explored. To fill up this lack of information, we present our detailed analysis of the dynamics of coupled BECs which demonstrates how to implement the whole set of single-qubit quantum logic gates. Our physical system consists of atoms in a magneto-optical trap, condensed in two different hyperfine levels which are coupled by a two-photon transition. First, we obtain an analytical solution for the time-dependent Schr\"odinger equation, written as an evolution operator. Then, we demonstrate that dynamics can be rewritten as successive applications of rotation operators, $\hat{R}_i$. Finally, we explore this general result in order to found the conditions over the physical parameters which control the dynamics of the system in such a way that corresponds to the action of six quantum logic operations: the gates NOT, Y, and Hadamard, which require transfer of population, and the phase gates $\hat{Z}$ ($\pi$), $\hat{S}$ ($\pi/2$), and $\hat{T}$ ($\pi/4$), which requires the inhibition of transfer of population using strong detuning.

Our system consists of condensed atoms in two different hyperfine levels labeled by $a$ and $b$. The interactions between the atoms are well described by assuming two-body collisions and the coupling of the two hyperfine levels is done via a two-photon transition. The second quantized Hamiltonian is written as $\left(\hbar=1\right)$~\cite{Milburn97,CiracGBEC98,Steel98,Villain99}
\begin{eqnarray}
\hat{H}&=&\tilde{\omega}_{a}\hat{n}_{a}+\tilde{\omega}_{b}\hat{n}_{b}+\gamma_{a}\hat{n}_{a}^{2}+\gamma_{b}\hat{n}_{b}^{2}
+2\gamma_{ab}\hat{n}_{a}\hat{n}_{b}\nonumber\\
&&-g\left(\hat{a}^{\dagger}\hat{b}e^{-\imath
\Delta t}+\hat{a}\hat{b}^{\dagger}e^{\imath \Delta t}\right),
\label{eq:h1}
\end{eqnarray}
where $\hat{n}_{a}=\hat{a}^{\dagger}\hat{a}$, $\hat{n}_{b}=\hat{b}^{\dagger}\hat{b}$, and we define the following physical parameters
\begin{eqnarray}
 \omega_{j}&=&\int d^{3}\vec{r}\Phi_{j}\left(\vec{r}\right)\left[ -\frac{1}{2m}\nabla^{2}+V_{j}\left(\vec{r}\right)\right]\Phi_{j}\left(\vec{r}\right),\\\nonumber
\gamma_{j}&=&\frac{4\pi A_{j}}{2m}\int d^{3}\vec{r} \Psi_{j}
^{4}\left(\vec{r}\right),\\\nonumber \gamma_{ab}&=&\frac{4\pi
A_{ab}}{2m}\int d^{3}\vec{r} \Psi_{a}
^{2}\left(\vec{r}\right)\Psi_{b}
^{2}\left(\vec{r}\right),\\\nonumber
g&=&\frac{\Omega}{2}\int
d^{3}\vec{r}
\Psi_{a}\left(\vec{r}\right)\Psi_{b}\left(\vec{r}\right).
\label{eq:par}
\end{eqnarray}
Here, $\omega_j$ describes the effect of harmonic potential over the condensed atoms, $\gamma_{j}$ depends on the scattering length $A_j$, describing collisions between the atoms condensed in the same hyperfine level $a$ or $b$, and $\gamma_{ab}$ is associated with the collisions between atoms in different levels. The coupling parameter $g$ takes into account the effect of a two-photon coupling between two different hyperfine levels with $\Omega$ being the coupling strength between the modes. This coupling could be resonant ($\Delta=0$) or not ($\Delta\neq0$). In equation (\ref{eq:h1}), we also use the auxiliary parameters $\tilde{\omega}_{(a,b)}=\omega_{(a,b)}-\gamma_{(a,b)}$.

The task here is to find the evolution operator $\hat{P}(t)$ which evolves some initial state. The evolved state, which is the solution of the Schr\"odinger equation associated with Hamiltonian (\ref{eq:h1}), can be written as
\begin{equation}
\ket{\Psi\left(t\right)}=\hat{P}(t)\ket{\Psi(0)}.
\label{eq:evolved}
\end{equation}
To find $\hat{P}(t)$, we use a technique which takes advantage of the properties of unitary transformations. We apply successively two transformations, being the first one, \[\hat{U}\left(t\right)=e^{\frac{-\imath \Delta t}{2}\left(\hat{n}_{a}-\hat{n}_{b}\right)},\]
which removes the explicit temporal dependence from original Hamiltonian (\ref{eq:h1}). The transformed Hamiltonian is non-diagonal in the Fock basis and nonlinear, once there is a term which depends on $\Delta\hat{n}^{2}$ operator. The coupling is given by the \textit{nonlinear parameter} $\Lambda$ defined as,
\begin{equation}
\Lambda\equiv\frac{1}{4}\left(\gamma_{a}+\gamma_{b}-2\gamma_{ab}\right).
\label{eq:lambda}
\end{equation}
In the case of Rubidium collisions it applies that $\Lambda=0$~\cite{Hall98a,Hall98b}, although the manipulation of this parameter is possible by using Feshbach resonances which affects collision between atoms from different hyperfine levels~\cite{Kaufman09,Erhard04}. Concerning the non-diagonal feature, a second transformation \[\hat{V}=e^{\frac{\xi}{2}\left(\hat{a}^{\dagger}\hat{b}-\hat{a}\hat{b}^{\dagger}\right)},\] is used to obtain the final form of a transformed Hamiltonian $\hat{H}^{V}=\hat{V}^{\dagger}\hat{H}^{U}\hat{V}$ given by
\begin{equation}
 \hat{H}^{V}=\omega_{0}\hat{N}+\gamma_{ab}\hat{N}^{2}+\left[\left(\omega_{1}+\omega_{2}\hat{N}\right)\cos\xi+g\sin\xi\right]\Delta\hat{n}.
\label{eq:hv}
\end{equation}
Here we define $ \omega_{0}=\left(\tilde{\omega}_{a}+\tilde{\omega}_{b}\right)/2$, $\omega_{1}=\left(\tilde{\omega}_{a}-\tilde{\omega}_{b}-\Delta\right)/2$,
$\omega_{2}=\left(\gamma_{a}-\gamma_{b}\right)/{2}$ and $\xi=\arctan\left(\frac{g}{\omega_{1}+\omega_{2}N}\right)$. $\hat{N} = \hat{n}_a+\hat{n}_b$ is the number operator, which accounts the total number of bosons in the system. Still, because $N$ is a conserved quantity, it will works as a c-number. After all this considerations, we obtain the final form for the evolved state~(\ref{eq:evolved}) which reads as
\[\ket{\Psi\left(t\right)}=\hat{U}\left(t\right)\hat{V}e^{-\imath\hat{H}^{V} t}\hat{V}^{\dagger}\ket{\Psi(0)}.\]
That means that $\hat{P}(t)$ is given by
\begin{equation}
\hat{P}(t)=\hat{U}\left(t\right)\hat{V}e^{-\imath\hat{H}^{V} t}\hat{V}^{\dagger}.
\label{eq:propagator}
\end{equation}
This evolution operator can be seen as a product of rotation operators, because of the connection between two quantum harmonic oscillators and pseudo-spin operators~\cite{Sakuraieng}, so  $\hat{J}_{x}=\hat{a}^{\dagger}\hat{b}+\hat{a}\hat{b}^{\dagger}$, $\hat{J}_{y}=-i\left(\hat{a}^{\dagger}\hat{b}-\hat{a}\hat{b}^{\dagger}\right)$, and $\hat{J}_{z}=\hat{a}^{\dagger}\hat{a}-\hat{b}^{\dagger}\hat{b}$. At the same time, the operators inside the general expression above are, in fact, rotation operators~\cite{Sakuraieng,Nielsen00}, once the operators $\hat{J}_i$ ($i=x,y,z$) are generators of rotations around $i$th axis. The final form for $\hat{P}(t)$ as a function of rotation operators $\hat{R}_{i}$ is
\begin{equation}
\hat{P}\left(\eta,\Delta,\varpi,\xi,t\right)=e^{-\imath\eta t} R_{z}\left(\Delta t\right)R_{y}\left(-\xi\right) R_{z}\left(\varpi t\right)R_{y}\left(\xi\right),
\label{eq:prot}
\end{equation}
where the auxiliary parameters are defined as
\begin{eqnarray}
\xi&=&\arctan\left(\frac{2g}{\Gamma-\Delta}\right),\nonumber\\
\varpi&=&\left(\Gamma-\Delta\right)\cos\xi+2g\sin\xi,\nonumber\\
\eta&=&\frac{1}{2}\left[\left(\omega_a+\omega_b\right)+\left(\gamma_a+\gamma_b\right)\left(N-1\right)\right]N.
\label{eq:etavarpi}
\end{eqnarray}
Here we define the \textit{frequency-scattering detuning} as
\begin{equation}
\Gamma\equiv\omega_{ab}+\left(\gamma_{a}-\gamma_{b}\right)\left(N-1\right),
\label{eq:gamma}
\end{equation}
being $\omega_{ab}=\omega_{a}-\omega_{b}$ the detuning between the effective frequency trap for each condensate. The parameter $\Gamma$ quantifies differences between the trap frequencies and collision parameters for each condensate. This quantity and the non-linear parameter, $\Lambda$, play important roles in the definition of the necessary conditions for the implementation of quantum logic gates.

After the obtention of the evolution operator, we proceed to exploit the possibilities of applications of coupled BECs within the context of quantum information processing. In the standard quantum computation approach, encoding information requires the preparation of an initial state, which is defined in the computational basis $\left\{\ket{0},\ket{1}\right\}$ as a coherent superposition with coefficients $\alpha$ and $\beta$. Definition of qubits using coupled BECs are found in the literature, particularly in the context of condensed atoms in a two-sites potential~\cite{Zeng02}. Our proposal is to encode the qubit using an atomic coherent state (ACS)~\cite{Arecchi72}, which can be created with hyperfine coupled BECs in two steps: i) the first one is the application of a pulse with a given duration to transfer the population from one condensate to another. For instance, if a $\pi/2$ pulse is applied over a BEC containing all condensed atoms, half of them will be in each hyperfine BEC after the pulse; and ii) the relative phase between the coupled BECs is created through the free evolution of the system~\cite{Hall98b,Oberthaler10,Riedel10}.
The ACS is defined as follows
\begin{eqnarray}
\ket{\theta,\phi}&=&\frac{1}{\sqrt{N!}}\left[\cos\left(\frac{\theta}{2}\right)
a^{\dagger}+\sin\left(\frac{\theta}{2}\right)e^{\imath\phi}
b^{\dagger}\right] ^{N}|0,0\rangle,\nonumber\\
&=&e^{\theta\left(\cos{\phi}\hat{J}_x+\sin{\phi}\hat{J}_y\right)}\ket{\frac{N}{2},-\frac{N}{2}},
\label{eq:acs}
\end{eqnarray}
which can be written as a quantum bit defined in the Bloch sphere as,
\begin{equation}
\ket{\theta,\phi}=\cos{\frac{\theta}{2}}\ket{0}+\sin{\frac{\theta}{2}}e^{i\phi}\ket{1}=\alpha\ket{0}+\beta\ket{1},
\label{eq:qbstate}
\end{equation}
with $\ket{0} = \ket{0,N}$ and $\ket{1}=\ket{N,0}$, $\theta$ and $\phi$ being the polar and azimuthal angles on Bloch sphere, respectively.
Here, the polar angle $\theta$ is related with the difference between atomic population of hyperfine levels while $\phi$ is the relative
phase between them. The state (\ref{eq:acs}) describes a particle with spin $N/2$, whose normalized mean values of the operators are given by $<Jx>/N=\sin{\theta}\cos{\phi}$, $<Jy>/N=\sin{\theta}\sin{\phi}$, and $<Jz>/N=\cos{\theta}$. ACS is analogous to the coherent state of the harmonic oscillator: it is the result of the application of a displacement operation over the ground state $\ket{N/2,-N/2}=\ket{1}$ and is a minimum uncertainty state which fulfills the Heisenberg angular-momentum uncertainty relation
$\meanval{\left(\Delta J_{x'}\right)^2}\meanval{\left(\Delta J_{y'}\right)^2}\geq\frac{1}{4}\left|\meanval{J_{z'}}\right|^2$, where the mean values are calculated in a rotated coordinates system so $z'$ is an axis which pass through the center of the ACS~\cite{Dowling94}.

Because $\hat{P}(t)$ is a succession of rotations, an initial state $\ket{\Psi(0)}$ prepared as an ACS evolves to another ACS with new coefficients $\alpha(t)$ and $\beta(t)$. To fulfill our goal of finding the experimental conditions for implementation of single-qubit quantum gates, we use the $2\times 2$ matrix representation of rotation operators $R_{z}$ and $R_{y}$~\cite{Nielsen00} and obtain the matrix form of $\hat{P}$ given by
\begin{equation}
\hat{P}\left(\eta,\Delta,\varpi,\xi,t\right)\rightarrow
e^{-\imath \eta t} \left(
\begin{array}{cc}
P_{11} & P_{12}\\
P_{21}& P_{22}
       \end{array}\right),
\label{eq:Pmatrix}
\end{equation}
where
\begin{eqnarray}
P_{11}&=&e^{-\imath\frac{\Delta t}{2}}e^{-\imath \frac{\varpi}{2}
t}\left(\cos^{2}\frac{\xi}{2} +e^{\imath \varpi
t}\sin^{2}\frac{\xi}{2}\right)\nonumber\\
P_{12}&=&2\imath \cos\frac{\xi}{2}\sin\frac{\xi}{2}
\sin\left(\frac{\varpi}{2} t\right) e^{-\imath\frac{\Delta t}{2}}\nonumber\\
P_{21}&=&2\imath \cos\frac{\xi}{2}\sin\frac{\xi}{2}
\sin\left(\frac{\varpi t}{2} \right) e^{\imath\frac{\Delta
t}{2}}\nonumber\\
P_{22}&=&e^{\imath\frac{\Delta t}{2}}e^{-\imath \frac{\varpi
t}{2}}\left( \sin^{2}\frac{\xi}{2}+e^{\imath \varpi
t}\cos^{2}\frac{\xi}{2}\right). \label{eq:pmatrizel}
\end{eqnarray}

Now we are able to find the necessary conditions for the realization of a specific single-qubit gate. This operations can be sorted in two sets, depending on the type of dynamics required for each gate. The first set, the \textit{transfer-population} gates, contains the NOT, Y, and Hadamard gates and requires transfer of population as well as changes on the relative phase. From the geometrical point of view, the evolution operator $\hat{P}(t)$ will change the polar angle and azimuthal angle of the qubit on Bloch sphere. The second set, the \textit{phase} gates, given by the Z, S, and T gates, requires the inhibition of transfer of population, together with the imprinting of the required relative phase \footnote{It is useful to remember the matrix form of gates NOT, $\hat{Y}$ and $\hat{H}$, which are written as
\begin{eqnarray}
\hat{X}\equiv\left(
\begin{array}{cc}
0 & 1\\
1 & 0
 \end{array}\right),\;& \hat{Y}\equiv\left(
\begin{array}{cc}
0 & -\imath\\
\imath & 0
 \end{array}\right),\;&\hat{H}\equiv\frac{1}{\sqrt{2}}\left(
\begin{array}{cc}
1 & 1\\
1 & -1
\end{array}\right)\nonumber,
\label{eq:gates1}
\end{eqnarray}
while the phase gates ($\hat{G}_p$) are given by the general expression
\[
\hat{G}_p=\left(
\begin{array}{cc}
1 & 0\\
0 & e^{\imath\varphi}
 \end{array}\right),
\label{eq:gates2}
\]
with the value of relative phase $\varphi$ corresponding to $\hat{G}_P=\hat{Z}$ ($\varphi=\pi$), $\hat{S}$ ($\varphi=\pi/2$), and $\hat{T}$ ($\varphi=\pi/4$) quantum gates, respectively.}.

To obtain the conditions for physical implementation, we proceed to compare the elements of evolution operator $\hat{P}$, given by equation (\ref{eq:pmatrizel}), with the
corresponding elements on the matrix representation for each gate. Let us consider the NOT quantum gate. Comparing both, the matrix form for this gate and equation (\ref{eq:Pmatrix}), the elements of $P$ must be given by $P_{11}=P_{22}=0$ and $P_{12}=P_{21}=1$. After some algebra, we found that the detuning must follow the rule $\Delta=\frac{2\pi}{t}$. In experiments, it is possible to control the detuning between the two-photon transition, $\Delta$, so we fix the gate evolution time as $t_{\rm{NOT}}=\frac{2\pi}{\Delta}$. Also, the values of the auxiliary parameters defined in equation (\ref{eq:etavarpi}) must fulfill the following conditions
\begin{eqnarray}
\xi=\arctan\left(\frac{2g}{\Gamma-\Delta}\right)&\equiv&\frac{\pi}{2},\nonumber\\
\varpi=\left(\Gamma-\Delta\right)\cos\xi+2g\sin\xi&\equiv&\frac{\Delta}{2},\nonumber\\
\eta=\frac{1}{2}\left[\left(\omega_a+\omega_b\right)+\left(\gamma_a+\gamma_b\right)\left(N-1\right)\right]N&\equiv&\frac{3\Delta}{8}.
\label{eq:parnot}
\end{eqnarray}
The solution of this coupled equations define two physical requirements: first, the value of the two-photon transition, $\Delta$, which must be set as $\Delta=\Delta_{\rm{G}}=4g$. Second, the frequency-scattering detuning, $\Gamma$, given by equation (\ref{eq:gamma}), must follow the condition $\Gamma=\Gamma_{\rm{G}}=4g$.

\begin{table}[ht]
\caption{\label{tab:gate1par}Evolution times and necessary conditions for the implementation of the transfer-population gates given by NOT, $\hat{Y}$, and $\hat{H}$. Here $g$ is the coupling parameter, $\Delta_{\rm{G}}$ is the detuning of the two-photon coupling, and $\Gamma_{\rm{G}}$ is the frequency-scattering detuning, as defined in equation (\ref{eq:gamma}).}
\begin{tabular}{cccc}
\hline
Gate& $t_{\rm{gate}}$&$\Delta_{\rm{G}}$&$\Gamma_{\rm{G}}$\\
\hline
NOT& $\frac{2\pi}{\Delta}$&$4g$       &$4g$\\
$\hat{Y}$   &$\frac{\pi}{\Delta}$        &$2g$       &$2g$\\
$\hat{H}$& $\frac{2\pi}{\Delta}$    &$\frac{8}{\sqrt{2}}g$&$-2g+\Delta$\\
\hline
\end{tabular}
\end{table}

Following a similar procedure, we found the values for $t_{\rm{gate}}$, $\Delta_{\rm{G}}$, and $\Gamma_{\rm{G}}$ for all transfer-population gates, summarized in table~\ref{tab:gate1par}. We note that the detuning of the two-photon transition, $\Delta_{\rm{G}}$, depends on coupling parameter $g$ with the rule $\Delta_{\rm{G}}=n^{\rm{G}}g$, with $n^{\rm{G}}$ being a different factor for each gate. The operation time is also a function of the coupling parameter $g$, which is responsible for the transfer of population, and all gates require a non-resonant two-photon coupling as well as specific non-zero values of the $\Gamma_{\rm{G}}$ parameter. This is explained because this two parameters ($\Delta_{\rm{G}}$ and $\Gamma_{\rm{G}}$) control the gain of relative phase between $\ket{0}$ and $\ket{1}$. For coupled BECs on a chip~\cite{Bohi2009}, it is possible to control $\omega_{ab}$ while $\gamma_{a}-\gamma_{b}$ is a fixed value. On the other hand, in the experimental setup in Ref.~\cite{Oberthaler10} the values of collision parameters $\gamma_{i}$ can be manipulated via Feschbach resonances, leaving $\omega_{ab}$ fixed.

Concerning the phase gates, once that the goal is to gain relative phase suppressing the transfer of population, we expect that the requirements for the implementation of this gate includes a detuned two-photon transition. For example, in order to obtain the gate $\hat{Z}$, it is necessary that the elements of propagator $P$ follow the conditions $P_{11}=1$, $P_{22}=-1$ with $P_{12}=P_{21}=0$. This requirements are fulfilled when the evolved time is given by $t_Z=\pi/2\Delta$. After some algebra, we obtain the conditions for the auxiliary parameters which are
\begin{eqnarray}
\xi=\arctan\left(\frac{2g}{\Gamma-\Delta}\right)&\equiv&\pi,\nonumber\\
\varpi=\left(\Gamma-\Delta\right)\cos\xi+2g\sin\xi&\equiv&3\Delta,\nonumber\\
\eta=\frac{1}{2}\left[\left(\omega_{a}+\omega_{b}\right)-\left(\gamma_{a}+\gamma_{b}\right)\right]N+\gamma_{ab}N^{2}&\equiv&\Delta.
\label{eq:parz}
\end{eqnarray}
There are two possible mathematical solutions for the coupled equations. One solution is given by the condition $g=0$, which is not interesting because describes a physical situation where the condensates are uncoupled. The second solution is obtained when $\Delta=\Delta_{\rm{G}}\gg2g/3$, which implies a strong detuning of the two-photon transition as expected. For $\hat{S}$ and $\hat{T}$, we obtain a similar condition for $\Delta_{\rm{G}}$ and, once a value of $\Delta_{\rm{G}}$ is fixed, a careful setting of quantity $\Gamma=\Gamma_{\rm{G}}$ is required in order to obtain the correct phase gain. With this considerations, we obtain the conditions for implementation of all phase gates that summarize in table~\ref{tab:gate2par}.
\begin{table}
\caption{\label{tab:gate2par}Evolution times and necessary conditions for the implementation of the quantum phase gates ($\hat{Z}$, $\hat{S}$, and $\hat{T}$). Here $g$ is the coupling parameter, $\Delta_{\rm{G}}$ is the detuning of the two-photon coupling, and $\Gamma_{\rm{G}}$ is the frequency-scattering detuning, as defined in equation (\ref{eq:gamma}).}
\begin{tabular}{cccc}
\hline
Gate&$t_{\rm{gate}}$&$\Delta_{\rm{G}}$   &$\Gamma_{\rm{G}}$\\
\hline
$\hat{Z}$ &$\frac{\pi}{2\Delta}$&$\gg \frac{2g}{3}$&$-2\Delta$\\
$\hat{S}$ &$\frac{3\pi}{2\Delta}$&$\gg 3g$&$\frac{\Delta}{3}$\\
$\hat{T}$ &$\frac{\pi}{2\Delta}$&$\gg 4g$ &$\frac{\Delta}{2}$\\
\hline
\end{tabular}
\end{table}

\begin{figure}[b]
\includegraphics[scale=1]{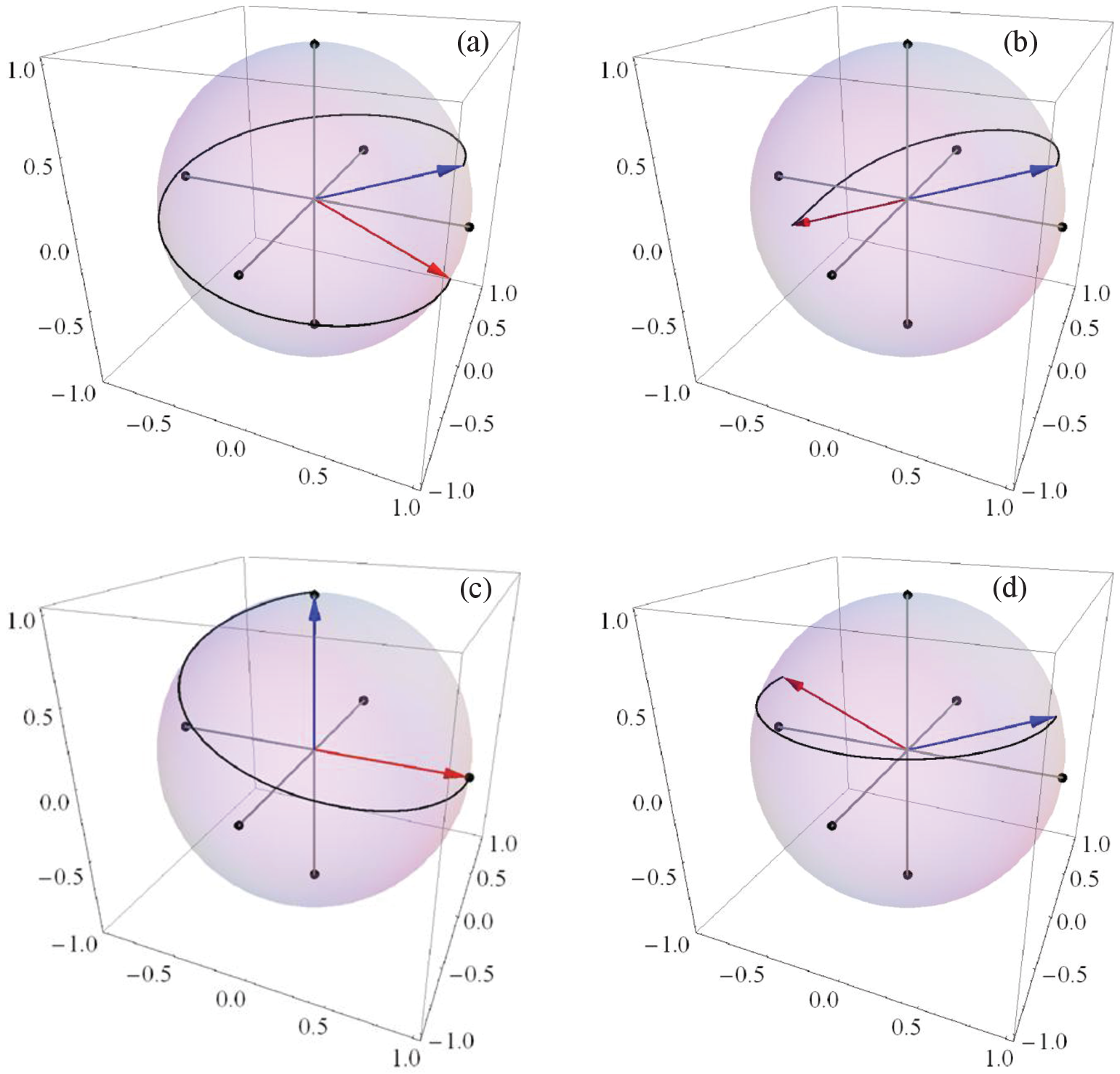}
\caption{Dynamics over the Bloch state considering the initial state $\ket{\Psi(0)}=\alpha\ket{0}+\beta\ket{1}$ under the action of (a) NOT, (b) $\hat{Y}$ and (c) $\hat{Z}$ considering $\alpha=\cos{\left(\frac{3\pi}{16}\right)}$ and $\beta=\sin{\left(\frac{3\pi}{16}\right)}$. (d) Action of the Hadamard gate over the initial state $\ket{\Psi(0)}=\ket{0}$ The blue arrow denote the initial state, while the red one denote the final state and $\left(x,y,z\right)\rightarrow\left(\frac{\meanval{\hat{J}_x}}{N},\frac{\meanval{\hat{J}_y}}{N},\frac{\meanval{\hat{J}_z}}{N}\right)$}.
\label{fig:xyzsgates}
\end{figure}

It is interesting to check the dynamics by calculating the mean values of the operators $\hat{J}_i$ ($i=x,y,z$) and follow their evolution over the Bloch sphere as shown in figure \ref{fig:xyzsgates}. First, we simulate the dynamics for  $\ket{\Psi(0)}=\alpha\ket{0}+\beta\ket{1}$ with $\alpha=\cos{\left(\frac{3\pi}{16}\right)}$, and $\beta=\sin{\left(\frac{3\pi}{16}\right)}$ under the conditions for the implementation of gates NOT, figure \ref{fig:xyzsgates}(a), $\hat{Y}$, figure \ref{fig:xyzsgates}(b), and the phase gate $\hat{Z}$, figure \ref{fig:xyzsgates}(c). For all cases, the initial state evolves to the target final state, in perfect agreement with the action of each quantum gate operation: $\ket{\Psi(t_{\rm{NOT}})}=\alpha\ket{1}+\beta\ket{0}$, for the NOT operation shown in figure \ref{fig:xyzsgates}(a), $\ket{\Psi(t_{\rm{Y}})}=\alpha\ket{1}-\beta\ket{0}$, for Y operation in figure \ref{fig:xyzsgates}(b), and $\ket{\Psi(t_{\rm{Z}})}=\alpha\ket{0}-\beta\ket{1}$ for Z operation in figure \ref{fig:xyzsgates}(c). The action of the Hadarmard gate is shown in figure \ref{fig:xyzsgates}(d) considering a different initial condition being $\ket{\Psi(0)}=\ket{0}$. As can be seen, the gate creates the superposition $\frac{\ket{0}+\ket{1}}{\sqrt{2}}$. We perform other simulations, not shown here, corroborating that initial state given by $\ket{1}$ evolves to the superposition $\frac{\ket{0}-\ket{1}}{\sqrt{2}}$, and the same gate takes both superpositions to the original initial states. From this results, our expectation is that the guidance provided by the tables~\ref{tab:gate1par}, and \ref{tab:gate2par}, together with a careful manipulation of physical parameters on coupled BECs, become a  practical path to follow to guarantee the implementation of all single-qubit quantum gates.

Concerning the feasibility of our proposal, the first task is to compare the evolution times for the quantum gates with the time scale of the typical decoherence processes for BECs. Let us consider the value of the coupling parameter for  experimental setup of Ref.~\cite{Bohi2009}, to calculate the evolution time $t_{\rm{gate}}$ for each gate. To estimate the value of coupling $g$ we use the reported value of $t_{\pi/2}=170$ $\mu$s of Ref.~\cite{Bohi2009}, associated with the application of a $\pi/2$ pulse, which means the two-photon coupling is $g=2\pi\times\frac{1}{4t_{\pi/2}}\cong2\pi\times1.5$kHz. Using the table~\ref{tab:gate1par}, we obtain that the evolution times associated with NOT, $\hat{Y}$, and $\hat{H}$ gates are given by $t_{\rm{NOT}}=t_{\hat{Y}}=0.16$ ms, and $t_{\hat{H}}=0.12$ ms, respectively. These values are significantly lower than the decoherence time, with value around 3 s~\cite{Treutlein04}. For the experimental setup of Ref.~\cite{Oberthaler10}, the value of the coupling parameter can be switched between zero and $g=2\pi\times 600$ Hz. Considering the highest value of $g$, we found the time scale for the gates as $t_{\rm{NOT}}=t_{\hat{Y}}=0.42$ ms, and $t_{\hat{H}}=0.29$ ms which is $10^{-4}$ times smaller than decoherence time. Because the phase gates require a detuned interaction ($\Delta_{\rm{G}}\gg g$), the time scale goes as $t_{\rm{gate}}\propto 1/\Delta_{\rm{G}}$.

The second and third tasks consist on check the effects of nonlinearity ($\Lambda\neq0$), and the sensibility of dynamics against changes on values of frequency-scatering detuning, $\Gamma$, and the two-photon detuning, $\Delta_{\rm{G}}$, as defined in tables~\ref{tab:gate1par} and~\ref{tab:gate2par}. Hao and Gu~\cite{Hao11} studied the dynamic associated with Hamiltonian~(\ref{eq:h1}), considering non-zero values for nonlinear condition. The nonlinearity, which depends on $\hat{J}^2_z$, would affect our qubit, once the ACS becomes a squeezed state~\cite{Oberthaler08} as time increases. This coupling also induces self-trapping, which might reduce the fidelity of quantum gates NOT, $\hat{Y}$ and $\hat{H}$. Changes on specific conditions for $\Gamma$ and $\Delta_{\rm{G}}$, could affect the fidelity of the implementation of quantum gates, because they are crucial for both, the necessary gain of relative phase and transfer of population.
To quantify the robustness of the gates, we perform simulations with the goal of solve the Schr\"{o}dinger equation to calculate the exact state $\ket{\Psi(t_{\rm{gate}})}$. We compare the match between the target gate state, $\ket{\Psi_{\rm{gate}}}$, and $\ket{\Psi\left(t_{\rm{gate}}\right)}$, calculating the fidelity defined as
\begin{equation}
\mathcal{F}=\left|\langle\Psi_{\rm{gate}}\ket{\Psi\left(t_{\rm{gate}}\right)}\right|^2.
\end{equation}

For our simulations we consider $N=1000$, the initial state being $\ket{\theta,\phi}=\ket{\pi/8,0}$, and values for physical parameters according with experimental setup of Ref.~\cite{Oberthaler10}. In order to quantify the effects of nonlinearity, we run the numerical simulations increasing the value of $\gamma_{ab}$ in order to obtain $\Lambda\neq 0$. Using a similar procedure, we quantify the effects of $\Gamma$ parameter defined for each gate on tables~\ref{tab:gate1par} and~\ref{tab:gate2par}, considering increasing values of $\omega_{ab}$ so $\Delta\Gamma=\left|\Gamma_s-\Gamma_{\rm{G}}\right|\neq 0$, being $\Gamma_s$ the value used at simulation and $\Gamma_{\rm{G}}$ the value in tables~\ref{tab:gate1par} and~\ref{tab:gate2par}. The effect of changes on $\Delta$ condition is quantified by considering $\delta\Delta=\left|\Delta_s-\Delta_{\rm{G}}\right|$, being $\Delta_s$ the value used at simulation.

\begin{figure}
\vspace{0.25cm}
\includegraphics[scale=0.75]{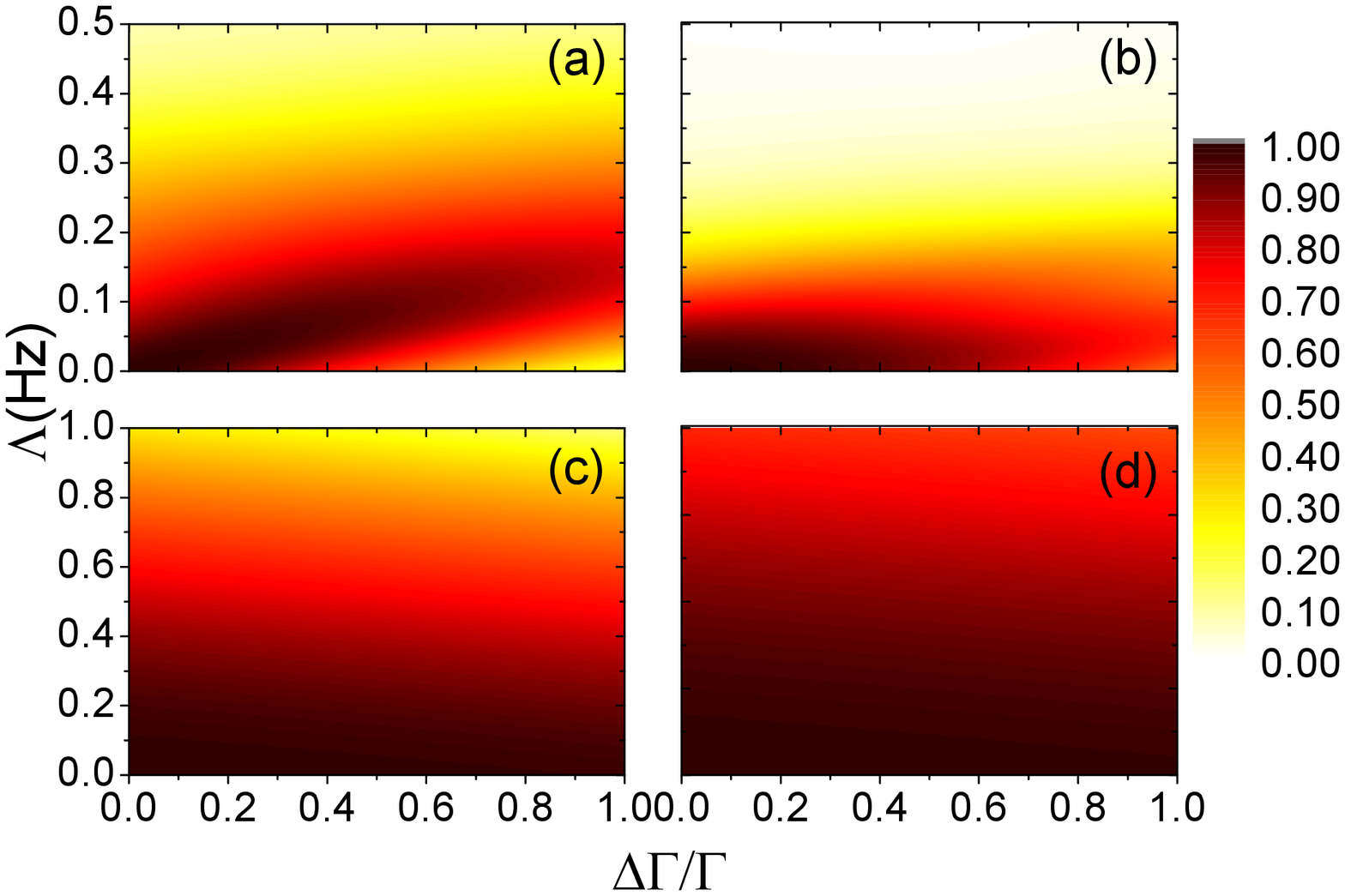}
\caption{(Color online) Fidelity of (a)$\hat{H}$, (b) $\hat{X}$ and $\hat{Y}$, (c) $\hat{S}$, and (d)$\hat{Z}$ gates as a function of $\Lambda$ and $\Delta\Gamma/\Gamma$}
\label{fig:fidxh}
\end{figure}

In figure \ref{fig:fidxh}, we plot our results of fidelity as a function of both, $\Lambda$ and $\Delta\Gamma/\Gamma$, considering four gates (a) $\hat{H}$, (b) $\hat{X}$ and $\hat{Y}$, (c)$\hat{S}$, and (d) $\hat{Z}$. For transfer-population gates, it is observed that the exact conditions for $\Lambda$ and $\Gamma_{\rm{G}}$ obtained by analytical calculations are not exclusive for the implementation of a gate with high fidelity: the dark region with $F=1$ extends to several choices for $\Lambda$ and $\Delta\Gamma/\Gamma$, with a linear relation between the two parameters for Hadamard gate, figure \ref{fig:fidxh}(a), and a less extended region for $\hat{X}$ and $\hat{Y}$ gates, figure \ref{fig:fidxh}(b). This two behaviors is understood checking the definition of $\Gamma$ parameter, equation (\ref{eq:gamma}): once this quantity depends on the collision parameters $\gamma_i$, for small values of $\Lambda$ the effect of nonlinearity can be compensated by modifying the frequency-scattering detuning, which provides an extended set of experimental values with high fidelity gates. In the case of phase gates, the effect of $\Delta\Gamma$ over the decay of fidelity is weaker than the effect of the nonlinearity parameter. This is explained by the fact that all phase gates do not require a specific value of two-photon detuning, being enough to set this parameter as significatively higher than coupling parameter $g$. Once that the condition for $\Gamma_{\rm{G}}$ depends directly of $\Delta$, it means that the variations around $\Gamma_{\rm{G}}$ ideal value on table~\ref{tab:gate2par} do not affect significatively the fidelity of gates.

\begin{figure}
\vspace{0.25cm}
\includegraphics[scale=1]{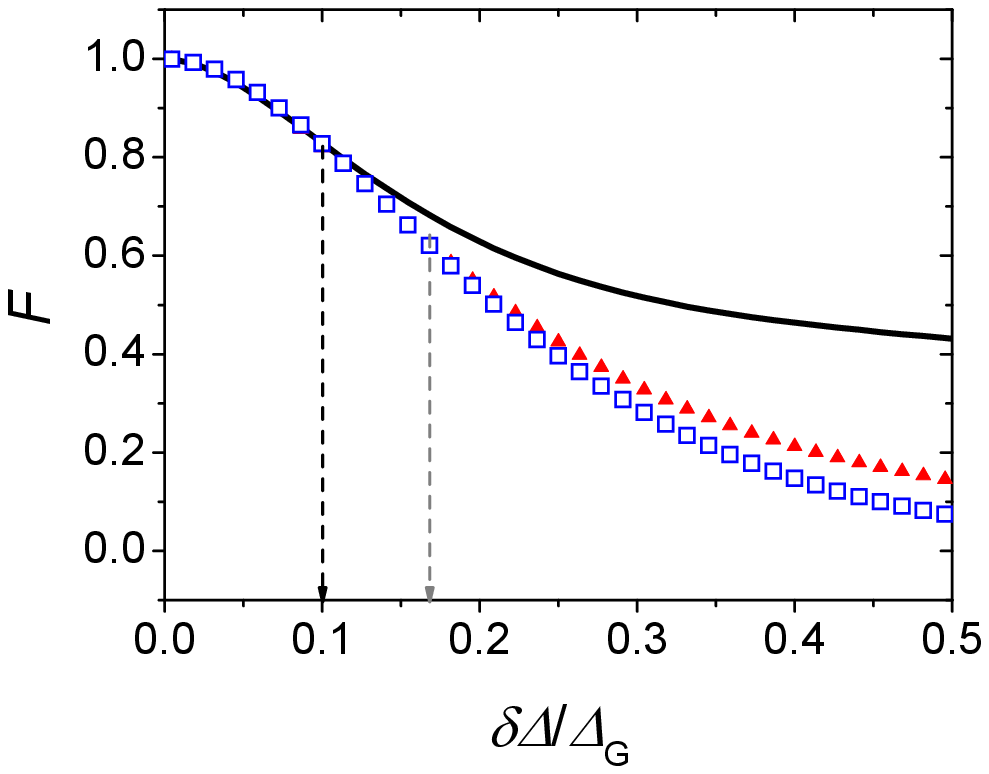}
\caption{(Color online) Fidelity behavior for $\hat{X}$ (black solid line), $\hat{Y}$ (red triangles) and $\hat{H}$ (blue open squares) gates as a function of $\Delta_{\rm{G}}$.}
\label{fig:fiddet}
\end{figure}

The last issue is to quantify the robustness of transfer-population gates out of the specific condition for detuning $\Delta_{\rm{G}}$ defined on table~\ref{tab:gate1par}. In figure \ref{fig:fiddet}, we show our numerical results for the three transfer-population gates. From our results, we observe that fidelity for all gates decays drastically when $\delta\Delta$ increases. Nevertheless, all three cases has the same behavior for small changes of $\Delta$ so $\delta\Delta/\Delta_{\rm{G}}\leq 0.1$. For this interval of values, the fidelity has high values with $F>0.8$, as indicated by the black dash arrow in figure \ref{fig:fiddet}. Additionally, the Hadamard and the $\hat{Y}$ gates continue with equal behavior until $\delta\Delta/\Delta_{\rm{G}}=0.17$, with $F>0.6$. After that value, the $\hat{X}$ gate fidelity decays slower than the other two gates. The last gates have the same behavior until $\delta\Delta/\Delta_{\rm{G}}=0.15$ (grey dash arrow in figure \ref{fig:fiddet}), with Hadamard being the less robust gate against small changes of the detuning value. From our simulations, we can conclude that fluctuations below $10\%$ can be considered small, because they are associated with high values of fidelity. Because $\Delta$ affects both, the gain of relative phase and transfer of population, the effect of imprecision on the exact value of two-photon detuning at $\Delta_{\rm{G}}$ increases when the gain of relative phase is mandatory on the implementation of the gate, which is the case for $\hat{Y}$ and $\hat{H}$.

Summarizing, we explore the properties of coupled Bose-Einstein condensates with the specific goal of implementation of controlled single-qubit quantum gates. We solve the Schr\"{o}dinger equation to explore the dynamics in order to obtain the necessary physical conditions for the implementation of two sets of gates: those which require transfer of population (NOT, $\hat{Y}$ and $\hat{H}$),  and the phase gates. We check the behavior of a Bloch vector defined using the mean values of pseudo-spin operators $\hat{J}_i$ with $i=x,y,z$ setting the necessary conditions for each quantum logic gate. We conclude that the careful experimental control of the different detunings of the system ($\Delta$, $\omega_a-\omega_b$ and $\gamma_a-\gamma_b$) leads to the implementation of high fidelity quantum gates. The feasibility of our proposal was explored through numerical simulations, considering the effects of nonlinearity and deviations from the required value for frequency-scattering condition, $\Gamma_{\rm{G}}$, as well as the two-photon detuning $\Delta_{\rm{G}}$. Our results show that it is possible to implement gates with high fidelity out of the ideal conditions. An open question is the scalability of this proposal. One option is the use of optical lattices~\cite{Blochrnat08,Folling07,Bloch03,Greiner02a,Greiner02b}, with each site containing this two-modes coupling Bose-Einstein condensates. An appealing path is to explore the possibility of the use of the quantum bus in Ref.~\cite{Byrnes15}, an extra coupling between two of these coupled BECs using a cavity. For both alternatives, it is necessary to evaluate how the extra coupling affects the gates which demands heavy numerical simulations. This problem are topic for future research.

This work was supported by the Brazilian National Institute of Science and Technology for Quantum Information (INCT-IQ), grant number 573658/2008-0, CAPES, grant number 552338/2011-7, FAPEMIG, grant number APQ-01768-14, and CNPq, grant number 552338/2011-7.
\section*{References}
\bibliographystyle{elsarticle-num}

%\bibliography{refluiz15}
\end{document}